\documentclass[aps,prl,reprint,superscriptaddress]{revtex4-2}
\usepackage{xcolor}
\usepackage{amsmath}
\usepackage{mathtools}
\usepackage{hyperref}
\usepackage{lineno}
\usepackage{capt-of}

\definecolor{bandpurple}{HTML}{CCCCFF}

\begin{document}

\title{First Measurement of Compton Scattering on $^3\text{He}$ at HI$\gamma$S} 


\author{J. Zhou}
\email{jingyi.zhou@duke.edu}
\affiliation{Department of Physics, Duke University, Durham, North Carolina 27708, USA}
\affiliation{Triangle Universities Nuclear Laboratory, Durham, North Carolina 27708, USA}

\author{E. Mancil}
\affiliation{Department of Physics, Duke University, Durham, North Carolina 27708, USA}
\affiliation{Triangle Universities Nuclear Laboratory, Durham, North Carolina 27708, USA}

\author{M. W. Ahmed}
\affiliation{Department of Mathematics and Physics, North Carolina Central University, Durham, North Carolina 27707, USA}
\affiliation{Triangle Universities Nuclear Laboratory, Durham, North Carolina 27708, USA}

\author{B. Crowe}
\affiliation{Department of Mathematics and Physics, North Carolina Central University, Durham, North Carolina 27707, USA}
\affiliation{Triangle Universities Nuclear Laboratory, Durham, North Carolina 27708, USA}


\author{G. Feldman}
\affiliation{Institute for Nuclear Studies, Department of Physics, The George Washington University, Washington DC 20052, USA}

\author{F. Q. L. Friesen}
\affiliation{Department of Physics, Duke University, Durham, North Carolina 27708, USA}
\affiliation{Triangle Universities Nuclear Laboratory, Durham, North Carolina 27708, USA}

\author{H. Gao}
\affiliation{Department of Physics, Duke University, Durham, North Carolina 27708, USA}
\affiliation{Triangle Universities Nuclear Laboratory, Durham, North Carolina 27708, USA}

\author{D. Godagama}
\affiliation{Department of Physics, Duke University, Durham, North Carolina 27708, USA}
\affiliation{Institute for Nuclear Studies, Department of Physics, The George Washington University, Washington DC 20052, USA}
\affiliation{Triangle Universities Nuclear Laboratory, Durham, North Carolina 27708, USA}

\author{H. W. Grießhammer}
\affiliation{Institute for Nuclear Studies, Department of Physics, The George Washington University, Washington DC 20052, USA}

\author{D. Hornidge}
\affiliation{Department of Physics, Mount Allison University 67 York St. Sackville, NB E4L 1E6, Canada}

\author{C. R. Howell}
\affiliation{Department of Physics, Duke University, Durham, North Carolina 27708, USA}
\affiliation{Triangle Universities Nuclear Laboratory, Durham, North Carolina 27708, USA}

\author{S. Jia}
\affiliation{Department of Physics, Duke University, Durham, North Carolina 27708, USA}

\author{S. Estupinan Jimenez}
\affiliation{Department of Physics and Astronomy, Montclair State University, Montclair, New Jersey 07043, USA}

\author{H. J. Karwowski}
\affiliation{University of North Carolina at Chapel Hill, Chapel Hill, North Carolina 27516, USA}
\affiliation{Triangle Universities Nuclear Laboratory, Durham, North Carolina 27708, USA}

\author{M. A. Kovash}
\affiliation{Department of Physics and Astronomy, University of Kentucky, Lexington, Kentucky 40506, USA}

\author{K. K. H. Leung}
\affiliation{Department of Physics and Astronomy, Montclair State University, Montclair, New Jersey 07043, USA}

\author{M. Lewis}
\affiliation{Institute for Nuclear Studies, Department of Physics, The George Washington University, Washington DC 20052, USA}

\author{Y. Liu}
\affiliation{Department of Physics, Duke University, Durham, North Carolina 27708, USA}
\affiliation{Triangle Universities Nuclear Laboratory, Durham, North Carolina 27708, USA}

\author{D. Markoff}
\affiliation{Department of Mathematics and Physics, North Carolina Central University, Durham, North Carolina 27707, USA}
\affiliation{Triangle Universities Nuclear Laboratory, Durham, North Carolina 27708, USA}

\author{C. Martin}
\affiliation{Department of Physics, Duke University, Durham, North Carolina 27708, USA}
\affiliation{Triangle Universities Nuclear Laboratory, Durham, North Carolina 27708, USA}


\author{J. McPherson}
\affiliation{Department of Chemistry and Physics, Longwood University, Farmville, VA 23909, USA}

\author{R. Mendez}
\affiliation{Department of Physics, Duke University, Durham, North Carolina 27708, USA}
\affiliation{Triangle Universities Nuclear Laboratory, Durham, North Carolina 27708, USA}

\author{S. F. Mikhailov}
\affiliation{Department of Physics, Duke University, Durham, North Carolina 27708, USA}
\affiliation{Triangle Universities Nuclear Laboratory, Durham, North Carolina 27708, USA}

\author{D. R. Phillips}
\affiliation{Department of Physics and Astronomy and Institute of Nuclear and Particle Physics, Ohio University, Athens, Ohio 45701, USA}

\author{M. H. Sikora}
\affiliation{Institute for Nuclear Studies, Department of Physics, The George Washington University, Washington DC 20052, USA}
\affiliation{Triangle Universities Nuclear Laboratory, Durham, North Carolina 27708, USA}

\author{N. Walton}
\affiliation{Department of Physics, Duke University, Durham, North Carolina 27708, USA}
\affiliation{Triangle Universities Nuclear Laboratory, Durham, North Carolina 27708, USA}

\author{J. Wang}
\affiliation{Department of Physics, Duke University, Durham, North Carolina 27708, USA}
\affiliation{Department of Physics, Universität Hamburg, Hamburg 20146, Germany}

\author{Y. K. Wu}
\affiliation{Department of Physics, Duke University, Durham, North Carolina 27708, USA}
\affiliation{Triangle Universities Nuclear Laboratory, Durham, North Carolina 27708, USA}

\author{B. Yu}
\affiliation{Department of Physics, Duke University, Durham, North Carolina 27708, USA}
\affiliation{Triangle Universities Nuclear Laboratory, Durham, North Carolina 27708, USA}

\author{Z. W. Zhao}
\affiliation{Department of Physics, Duke University, Durham, North Carolina 27708, USA}
\affiliation{Triangle Universities Nuclear Laboratory, Durham, North Carolina 27708, USA}

\collaboration{Compton@HI$\gamma$S Collaboration}
\noaffiliation

\date{\today}

\begin{abstract}

The world's first Compton scattering measurements on a $^3$He target were performed to probe neutron electromagnetic polarizabilities. Differential cross sections were measured at laboratory angles of $40^\circ$, $55^\circ$, $90^\circ$, $115^\circ$, $150^\circ$, using circularly polarized, quasi-monoenergetic photon beams at weighted mean energy values of 61.5 and 97.1 MeV from the High Intensity $\gamma$-ray Source at the Triangle Universities Nuclear Laboratory. Comparisons with chiral effective field theory show good agreement at 97.1 MeV but over-prediction at 61.5 MeV, motivating further theoretical development.



\end{abstract}


\maketitle

Nucleons are the building blocks of atomic nuclei, which account for over 99\% of visible matter by mass in the Universe. They serve as one of the most important windows into how quantum chromodynamics (QCD) works in the non-perturbative regime, where different theoretical methods have been established, such as lattice QCD, effective field theory, and QCD-motivated models.  To fully test the theory, various aspects of nucleon structure must be investigated, with the polarizabilities representing one important set of properties. The nucleon electric and magnetic polarizabilities ($\alpha_{E1}$ and $\beta_{M1}$)~\cite{Schumacher:2019ikn,Griesshammer12} encode the response of the nucleon's constituents to an applied electromagnetic field, and thus provide detailed information about the internal structure and interactions of its active degrees of freedom.

Precision measurements of the nucleon polarizabilities are among the most direct and stringent experimental tests of chiral symmetry breaking and pion dynamics in low-energy QCD. These quantities can be calculated directly using lattice QCD, which employs quarks and gluons as fundamental degrees of freedom~\cite{detmold2006electromagnetic,detmold2010extracting,Bignell:2020xkf,Hall:2013dva}. This approach has seen significant recent progress~\cite{Lujan:2016ffj,Engelhardt:2011qq,Saenz:2020yxy,Saenz:2020yxy,Lee:2024dgf}. Notably, the latest lattice predictions of $\alpha_{E1}$ are consistent with experimental results~\cite{wang2024nucleon}, establishing a strong foundation for computing other polarizabilities. Complementarily, in the low-energy regime, nucleon polarizabilities are described by chiral effective field theory ($\chi$EFT), which utilizes pions and nucleons as the effective degrees of freedom. Consequently, extractions of the nucleon polarizabilities via $\chi$EFT provide a critical test of lattice QCD, and in turn, of $\chi$EFT as a bridge between QCD and data. These theoretical methods describe nucleon properties with different degrees of freedom, and thereby provide a pathway to a comprehensive understanding of QCD.


Compton scattering is currently the premier experimental tool for accessing nucleon polarizabilities, as photons can provide an electromagnetic field strong enough to induce measurable responses~\cite{drechsel2003dispersion,Hagelstein15, Hagelstein:2020vog,Griesshammer12}.
The scalar polarizabilities $\alpha_{E1}$ and $\beta_{M1}$ appear as structure constants in the Compton scattering amplitude and are encoded in physical observables such as differential cross sections and beam/target asymmetries. 
The differential cross sections at forward angles are sensitive to the sum of the scalar polarizabilities, $\alpha_{E1}+\beta_{M1}$, while those at backward angles are sensitive to their difference, $\alpha_{E1}-\beta_{M1}$~\cite{maximon1989scattering}. In addition, $\alpha_{E1}+\beta_{M1}$ can be independently constrained by the Baldin sum rule (BSR)~\cite{baldin1960polarizability}. Beyond this angular dependence, the overall sensitivity of the cross sections to the polarizabilities grows quadratically as the incident photon beam energy increases.
Theoretical approaches including $\chi$EFT and dispersion relations are used to extract the polarizabilities from data, see e.g.~\cite{Hagelstein15, Schumacher:2019ikn, Griesshammer12}.

The recommended values from the Particle Data Group (PDG) in 2024~\cite{ParticleDataGroup:2024cfk} for the nucleon electromagnetic polarizabilities are:
\begin{equation}\label{eq:PDG_2025}
\begin{aligned}
    \alpha^p_{E1} &= (11.5 \pm0.4),\quad  \beta^p_{M1} = (2.31 \pm 0.29),\\
    \alpha^n_{E1} &= (11.8 \pm 1.1), \quad    \beta^n_{M1} = (3.7 \pm 1.2),
\end{aligned}
\end{equation}
where the polarizabilities given here and throughout this Letter are in units of $10^{-4}~\text{fm}^3$. These values are obtained by taking weighted averages of selected extractions via Compton scattering and other methods. The uncertainties include both experimental and theoretical uncertainties, adjusted by a scale factor.

While scalar polarizabilities of the proton are relatively well-determined, studies of the neutron are experimentally more challenging due to its charge neutrality and finite lifetime. Consequently, experiments must rely on light nuclear targets as ``effective'' neutron targets, making extraction of the neutron polarizabilities more complicated. 
Despite these challenges, investigating the neutron is as vital as studying the proton for a complete understanding of nucleon structure in QCD. 
Moreover, differences between proton and neutron polarizabilities encode isospin and chiral symmetry-breaking effects, offering a crucial test of the effective degrees of freedom in the low-energy region of $\chi$EFT.  
In particular, determining whether $\beta^n_{M1}>\beta^p_{M1}$ provides crucial information regarding sub-leading effects in the magnetic response of the nucleon, and the precise measurement of $\alpha^p_{E1} - \alpha^n_{E1}$ provides a stringent experimental test of the Cottingham Sum Rule prediction~\cite{Gasser:2015dwa} ($\alpha^p_{E1} - \alpha^n_{E1} = -1.7 \pm 0.4$) for the electromagnetic self-energy correction to the proton-neutron mass difference~\cite{long2025scattering}. 

Previous extractions of the neutron polarizabilities have primarily relied on measurements using deuterium targets, and representative results from elastic Compton scattering are summarized in Table~\ref{tab:previous_extractions}. 
In the table, Global analysis 1 evaluated data from three earlier experiments~\cite{lucas1994compton,Hornidge:1999xs,Lundin:2002jy} using $\chi$EFT~\cite{Griesshammer12,Myers:2015aba}. The work by Myers \textit{et al.} represents the most precise extraction from a single measurement, conducted at MAX-lab and evaluated using $\chi$EFT~\cite{Myers:2015aba}. Global analysis 2~\cite{COMPTONMAX-lab:2014cve, Myers:2015aba} combines the Myers \textit{et al.} measurement with the three earlier datasets, evaluated within the $\chi$EFT framework. As shown, while the neutron polarizabilities extracted using $\chi$EFT are consistent within their uncertainties, these bounds remain too large to clearly resolve the underlying dynamics or provide a rigorous test of the theory. As shown in Table~\ref{tab:previous_extractions}, the current uncertainties of the neutron polarizabilities are still dominated by the experimental uncertainties, which motivate subsequent high-precision measurements. 

Over the past decade, a series of Compton-scattering experiments have been conducted at the High Intensity $\gamma$-Ray Source (HI$\gamma$S) facility~\cite{Weller:2009zza} with hydrogen~\cite{Li:2022vnz}, deuterium~\cite{Godagama:2022aiy}, $^4$He~\cite{Li:2019irp}, and $^6$Li ~\cite{Myers:2012xw, myers2014compton} targets aiming to provide the most precise determination of nucleon electromagnetic polarizabilities.

\label{eq:Myers_neutron_combined}

\begin{table*}
\caption{\label{tab:previous_extractions}{Summary of extracted neutron polarizabilities from previous measurements of elastic Compton scattering on the deuteron. Values are given in units of $10^{-4}~\text{fm}^3$, and all extractions are constrained by the BSR. The quoted `exp' uncertainties represent the combined statistical and angle-dependent systematic uncertainties, added in quadrature.} 
}
\begin{ruledtabular}
\begin{tabular}{lll}
Analysis & $\alpha^n_{E1}$ & $\beta^n_{M1}$ \\ \hline
Global analysis 1~\cite{lucas1994compton,Hornidge:1999xs,Lundin:2002jy} & $11.1 \pm 1.8(\text{exp}) \pm 0.2(\text{BSR}) \pm 0.8(\text{th})$ & $4.1 \mp 1.8(\text{exp}) \mp 0.2(\text{BSR}) \mp 0.8(\text{th})$  \\
Myers \textit{et al.}~\cite{Myers:2015aba} & $13.66 \pm 1.6(\text{exp}) \pm 0.2(\text{BSR}) \pm 0.8(\text{th})$ & $1.65 \mp 1.6(\text{exp}) \mp 0.2(\text{BSR}) \mp 0.8(\text{th})$ \\
Global analysis 2~\cite{COMPTONMAX-lab:2014cve, Myers:2015aba,lucas1994compton,Hornidge:1999xs,Lundin:2002jy}. 
& $11.55 \pm 1.25(\text{exp}) \pm 0.2(\text{BSR}) \pm 0.8(\text{th})$ & $3.65 \mp 1.25(\text{exp}) \mp 0.2(\text{BSR}) \mp 0.8(\text{th})$ \\
\end{tabular}
\end{ruledtabular}
\end{table*}


This Letter reports the world’s first measurement of elastic Compton scattering on a $^3 \rm He$ target and the first absolute differential cross-section data for this process. Two decades ago, the first theoretical calculation of elastic Compton scattering on $^3$He identified the reaction as a highly promising candidate for extracting the neutron electromagnetic polarizabilities~\cite{choudhury2007investigating}. Due to technical challenges, such measurements only became possible recently with advances on the experimental front. With these developments, the experimental side has not only caught up, but also perhaps surpassed theory. The measurements were performed using quasi-monoenergetic, circularly polarized photon beams with intensity-weighted mean incident photon energies of 61.5 and 97.1 MeV.
The results are compared with the latest $\chi$EFT predictions~\cite{Margaryan:2018opu}, which mapped out the sensitivity of the Compton scattering differential cross section to neutron electromagnetic polarizabilities at energies between 50 and 120 MeV.
While $\chi$EFT calculations have demonstrated good consistency with data for the deuteron and $^4$He~\cite{long2025scattering}, Compton scattering on these nuclei is only sensitive to the isoscalar (average) nucleon polarizabilities. In contrast, $^3$He provides access to a distinct linear combination ($2p+n$). Consequently, this first-ever comparison with $^3$He data serves as a crucial cross-check of the $\chi$EFT formalism and provides a rigorous method for reducing systematic uncertainties in the extracted polarizabilities. 
The experiment was carried out at the HI$\gamma$S facility, which produced nearly 100\% circularly polarized photon beams through Compton backscattering of free-electron laser light from relativistic electron bunches circulating in a storage ring~\cite{Weller:2009zza}. The experimental configuration is shown in Fig.~\ref{HIGS_production_config}. The photon beams were generated at a rate of $\sim$5.6~MHz, providing beam bunches separated by $\sim$179~ns. This timing structure allowed for the effective identification of beam-correlated events and the rejection of cosmic-ray and other accidental (i.e., random in time) backgrounds in the detection system. 
To ensure a well-defined beam profile and facilitate the separation of the inelastic $^3$He channels from the elastic Compton scattering peak, the beam was collimated for the 61.5 MeV (97.1 MeV) measurements by a 16.0 (12.7) mm diameter lead collimator located upstream of the target area. 
These yielded the intensity-weighted mean energy of 61.5~MeV (97.1~MeV) with energy spread of 6.3\% (4.7\%)~(FWHM). Determination of the beam profiles can be found in Refs.~\cite{em2026,jz2026}. 
The collimated photon beam intensity was monitored using a newly built low-efficiency 4-paddle system~\cite{em2026,jz2026,Friesen_HIgS_Flux}. The average on-target photon beam intensities were $1.4\times 10^7~\gamma/$s ($3\times 10^6~\gamma/$s) at 61.5~MeV (97.1~MeV), and the systematic uncertainties in the intensity measurements were 2\% (3\%), respectively (see Refs.~\cite{em2026,jz2026} for details). 

The photons were scattered from a cryogenic target containing liquid $^3$He at 1.76~K within a 20-cm-long, truncated-cone-shaped cell made with 0.1-mm-thick Kapton sheets~\cite{kendellen2016cryogenic, Leung:DNP2024, 3He_Target}. The data were collected with the target cell both full and empty. The full-cell configuration yielded an areal target thickness of $(3.17\pm0.03)\times10^{23}$~nuclei/cm$^2$, while the empty-cell measurements were used to subtract background contributions from the target structure.
The 350 STP-liter inventory of $^3$He contains a small fraction of $^4$He. The $^4$He-to-$^3$He ratio in the target liquid was determined to be (1.9$\pm$0.3)\% based on residual gas analyzer measurements.

\begin{figure}[tbhp]
\includegraphics[width=3.0in]{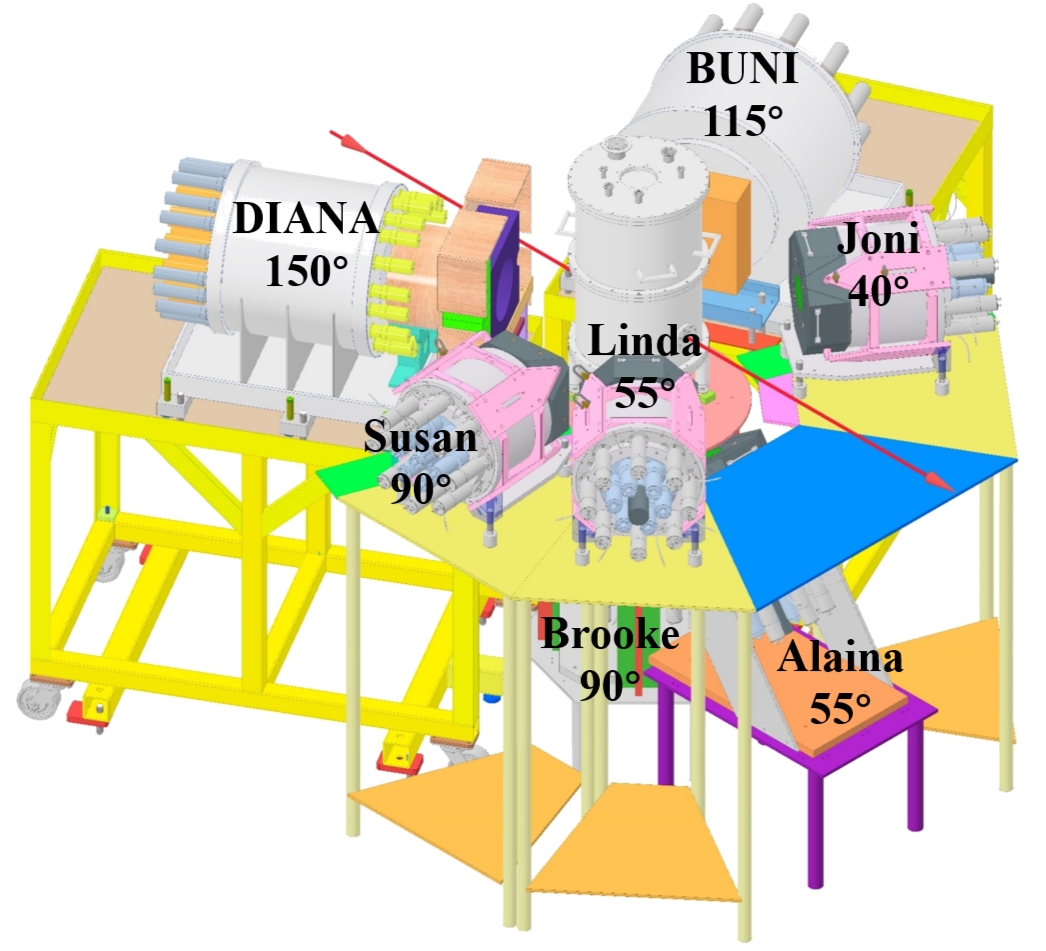}
\caption{Detector and target setup in the target room. The arrow indicates the direction of the photon beam. }\label{detector_setup}
\end{figure}

Photons scattered from the target were detected by an array of seven NaI(Tl) detectors, as shown in Fig.~\ref{detector_setup}.
Each detector consisted of a cylindrical NaI(Tl) core surrounded by a NaI(Tl) annulus and several plastic scintillator veto paddles for rejecting cosmic rays, the dominant background in the experiment. 
Two large-volume detectors, BUNI~\cite{Miller:1988gh} and DIANA~\cite{shoniyozov2016elastic}, were positioned at $115^\circ$ and $150^\circ$ in the laboratory frame, respectively. This arrangement leverages their superior energy resolution and background rejection capabilities for measurements at the backward angles, where differential cross sections have greater sensitivity to the neutron polarizability difference.
The remaining five medium-sized detectors from the HI$\gamma$S NaI Detector Array (HINDA)~\cite{henshaw2010investigation} were distributed across the forward and medium angles. Three of these were aligned in the horizontal plane and positioned at laboratory angles of $40^\circ, 55^\circ$, and $ 90^\circ$, while the other two were positioned in the vertical plane at $55^\circ$ and $ 90^\circ$. The geometry of the experimental apparatus (see more details in Refs.~\cite{em2026,jz2026}) was surveyed and incorporated into a GEANT4~\cite{GEANT4:2002zbu} Monte Carlo simulation, which was essential for the subsequent data analysis.

Detector signals and the beam time structure were recorded using ten 14-bit CAEN V1730(S) digitizers with a sampling rate of 500 MHz and a digitization window of 2~$\mu$s. The digitized waveforms were processed via baseline subtraction and pulse integration to extract the energy deposition for each detector segment for each event. A software-based constant fraction discriminator algorithm determined the detector trigger time, and the time-of-flight (ToF) for each event was defined as the difference between this trigger and arrival of the subsequent beam pulse. This optimized algorithm achieved a ToF precision of 0.5–1.0~ns (1$\sigma$)~\cite{em2026, jz2026}, the highest to date for Compton scattering at HI$\gamma$S, which was critical for maximizing the signal-to-background ratio and controlling systematics. Furthermore, gain corrections for individual detector segments were applied to improve energy resolution, enhancing the separation of inelastic scattering channels. 

\begin{figure}[tbhp]
\includegraphics[width=\linewidth]{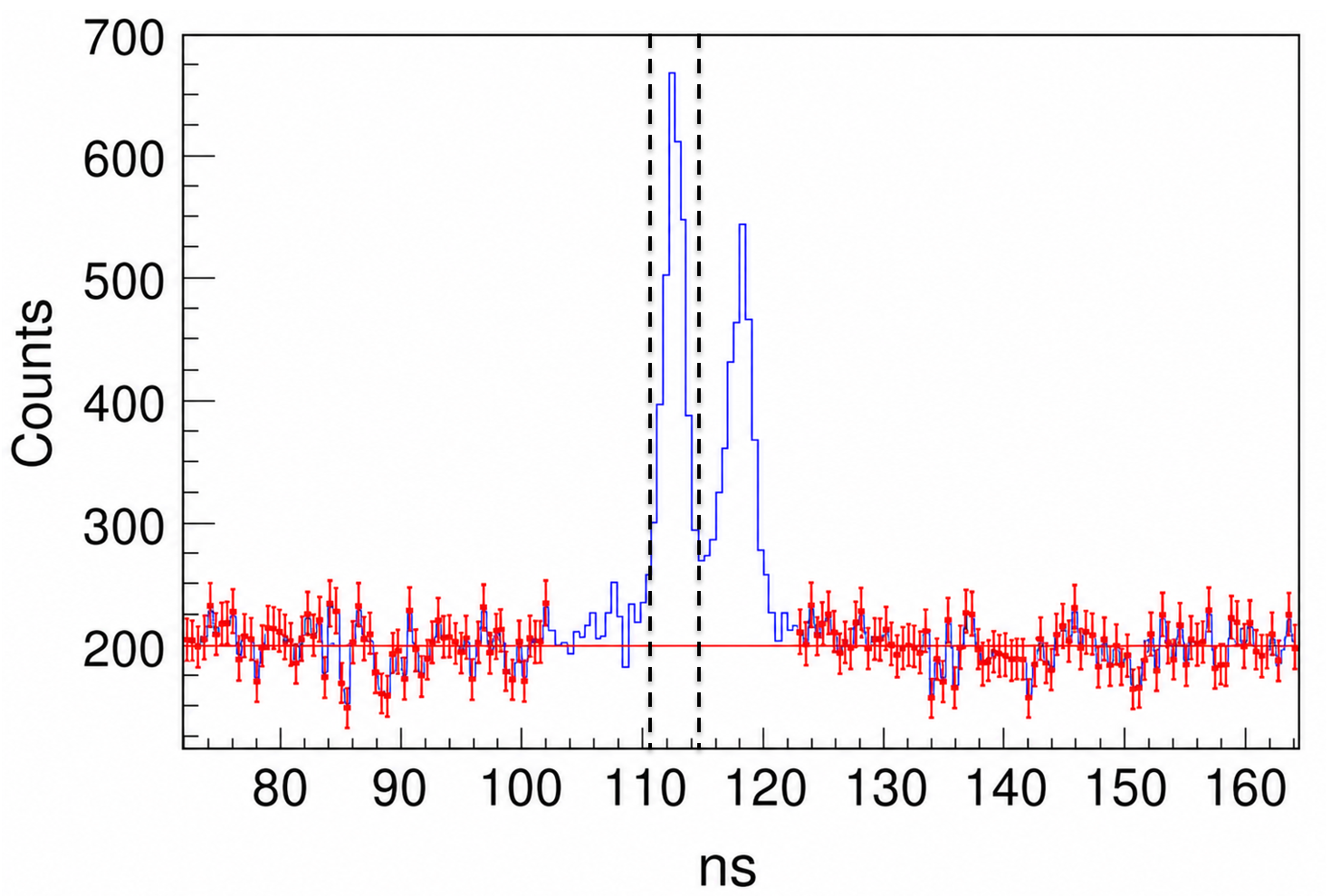}
\caption{Representative ToF spectrum. The region between the two black dashed lines defines the prompt ToF window, while the regions marked by the red data points correspond to the random events outside the prompt ToF windows. The peak on the right side of the prompt window originates from incident photons scattering off beamline materials located upstream of the target. }\label{ToF_Spectrum_DIANA}
\end{figure}
\begin{figure*}[tbhp]
\includegraphics[width=5.5in]{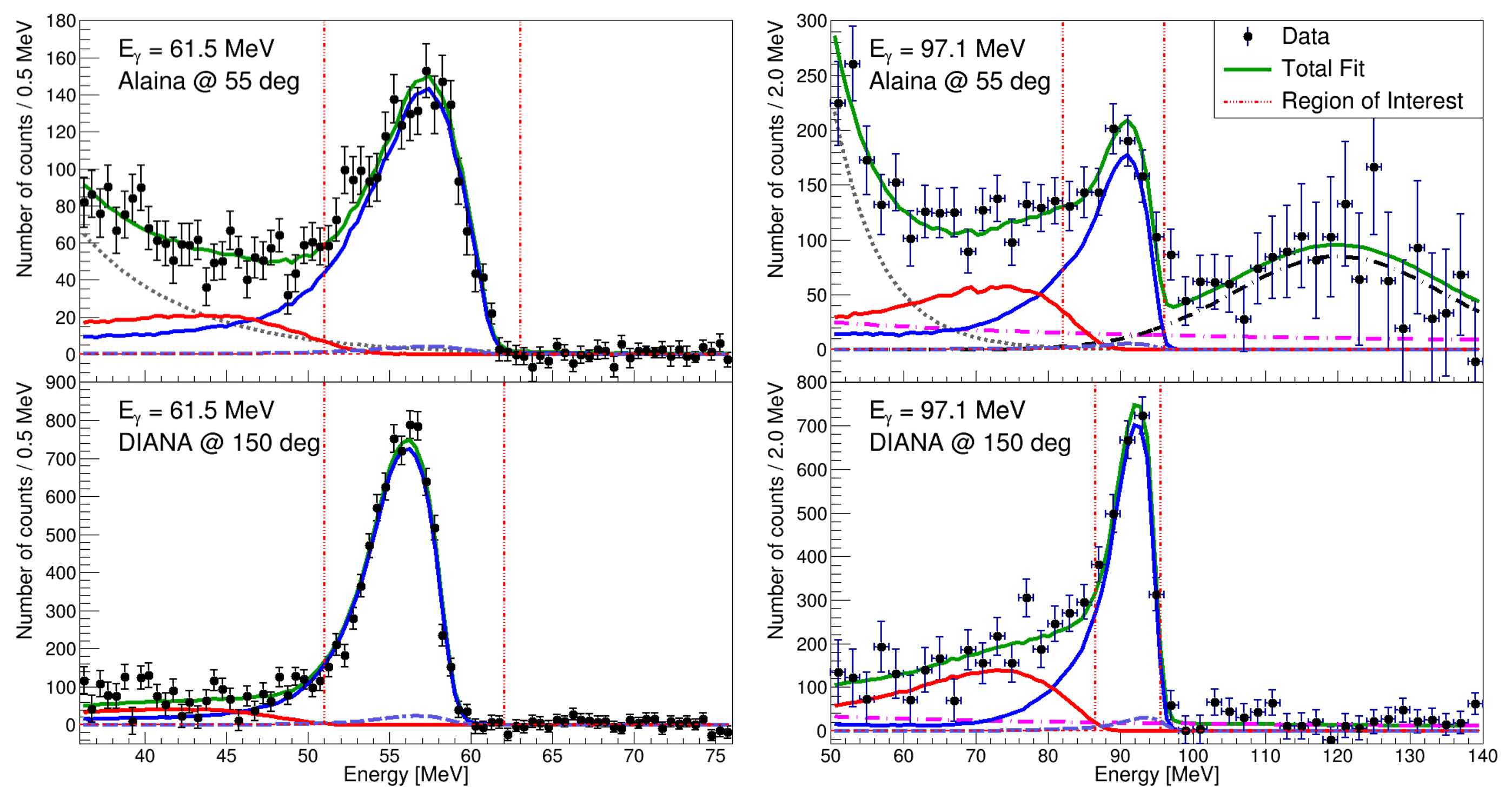}
\caption{Representative empty-target-subtracted spectra and typical fit results at 61.5 MeV (left), 97.1 MeV (right), 55 deg (top), 150 deg (bottom). Error bars on the data points represent statistical uncertainties. The vertical dotted lines indicate the optimized ROI. The total fit (solid green curve) comprises the GEANT4-simulated $^3$He elastic signal response function (solid blue), the $^4$He elastic background (dashed purple), the $^3$He quasi-elastic background (solid red), and other residual backgrounds. These include the atomic background (dotted black) at both energies, along with the bremsstrahlung (dot-dashed magenta) and pion photoproduction (dot-dashed black) backgrounds present at 97.1 MeV. The horizontal thin dotted red lines are zero-count lines.
}\label{fig:LineShapeFit}
\end{figure*}

To extract the signal yield, energy cuts on the annuli and plastic veto paddles were first applied to reject the cosmic-ray background. The pulsed nature of the photon beam produced a ToF spectrum with a clear prompt peak of beam-correlated events above a flat random background, as shown in Fig.~\ref{ToF_Spectrum_DIANA}. ToF cuts were then used to select the Compton scattering signal, reject out-of-time backgrounds, and further suppress cosmic rays. The signal efficiencies of these cuts were evaluated and applied as correction factors to the final differential cross sections. Together, these cuts rejected over 99\% of the cosmic rays within the energy region of interest (ROI). As shown in Fig.~\ref{fig:LineShapeFit}, ROI is defined to capture the majority of the signal. The boundaries of this region were optimized for each detector to maximize the signal-to-background ratio and minimize the combined statistical and systematic uncertainties of the extracted signal yield. 
Next, residual accidental events that remain in the prompt ToF windows were removed by subtracting the time-normalized energy spectrum of a random ToF window from that of the prompt window. The procedure above was applied to both full- and empty-target datasets. Finally, the empty-target data, normalized by the integrated beam photons, were subtracted from the full-target data. 

The last step of the signal yield extraction involved fitting the empty-target-subtracted spectrum using response functions generated by a GEANT4~\cite{GEANT4:2002zbu} simulation and assumed analytic functional forms. The simulation accounts for the beam energy profile, intrinsic detector resolution, extended target geometry, finite detector acceptances, and photon attenuation in the target and surrounding materials. This simulation package has been validated through comparisons with detector in-beam and cosmic-ray measurements. 
The modeled response functions include the elastic $^3$He Compton scattering signal and two major residual backgrounds: quasi-elastic $^3$He scattering (breakup) and elastic scattering from the small $^4$He contamination in the target. Reaction kinematics were also explicitly incorporated~\cite{jz2026}. Representative fits are shown in Fig.~\ref{fig:LineShapeFit}. 
For the forward-angle detectors, a low-energy atomic background was prominent and modeled with an exponential function. At 97.1 MeV, accelerator conditions introduced bremsstrahlung photons reaching up to 1 GeV, producing an additional background. This background was characterized by a continuous distribution with yield inversely proportional to energy. These high-energy photons also generated a $\pi^0$ background at forward angles, where one of the two $\pi^0$ decay photons was detected. This background source was confirmed via coincidence event searches~\cite{jz2026} and was described using a Gaussian distribution. Finally, the fitted residual backgrounds were subtracted from the spectra to obtain the net $^3$He signal yield for each detector.

The differential cross section extracted from each detector is calculated as follows:
\begin{equation}
\label{eq:diffxs}
\frac{d\sigma}{d\Omega}
=
\frac{Y_{\mathrm{ROI}} \, f_{\mathrm{bcc}}}
     {N_{\gamma} \, t \, f_{\mathrm{in}} \, \epsilon \, A_{\mathrm{eff}}} \, .
\end{equation}
where $Y_\text{ROI}$ is the integrated signal yield within the ROI. Here, the normalization factors common to all detectors include the total number of incident photons ($N_{\gamma}$), the target thickness ($t$), and the incident photon attenuation correction ($f_\text{in}$). The detector-dependent factors include the event selection efficiency ($\epsilon$), the bin-center correction factor ($f_\text{bcc}$), and the effective acceptance ($A_\text{eff}$); both $f_\text{bcc}$ and $A_\text{eff}$ were determined via simulation.

Systematics are categorized into two types: overall normalization uncertainties and point-to-point uncertainties. 
The overall normalization uncertainties include the total number of incident photons and the target thickness, as listed in Table \ref{tab:systematics}.
These two sources of uncertainties were summed in quadrature to determine the total normalization systematic uncertainty. Because this type of uncertainty affects the absolute magnitude of the differential cross sections but not their angular distributions, it does not impact the extraction of the neutron electromagnetic polarizabilities difference ($\alpha^{n}_{E1}-\beta^{n}_{M1}$). Point-to-point uncertainties arise from detector-specific analysis methods, such as event selection cuts or detector response modeling. 
Primary sources are listed in Table~\ref{tab:systematics}, which include uncertainties in the beam energy profile, the effective acceptance, the event selection cut efficiencies, and the fitting procedure~\cite{em2026,jz2026}.
Notably, fitting uncertainties increased at 97.1~MeV due to the presence of additional backgrounds. These point-to-point uncertainties were summed in quadrature to determine the total point-to-point systematic uncertainty for each data point, and are given in Table~\ref{tab:cross_sections}. A detailed discussion
of the analysis and uncertainties can be found in Refs.~\cite{em2026,jz2026}.

\begin{table}[!htbp]
\caption{\label{tab:systematics}
Summary of overall normalization and point-to-point systematic uncertainties at 61.5 and 97.1~MeV.} 
\begin{ruledtabular}
\begin{tabular}{lcc}
Source & 61.5~MeV & 97.1~MeV \\
\colrule
\multicolumn{3}{c}{\textit{Overall normalization}} \\
\colrule
Incident photons & 2\% & 3\% \\
Target thickness & 1\% & 1\% \\
\colrule
\multicolumn{3}{c}{\textit{Point-to-point}} \\
\colrule
Beam energy profile & 0.1\%--0.7\% & 0.1\%--0.7\% \\
Effective acceptance & 1.0\%--1.5\% & 1.6\%--1.9\% \\
Event selection & 0.3\%--2.8\% & 1.2\%--3.1\% \\
Fitting & 0.9\%--2.5\% & 0.8\%--7.4\% \\
\end{tabular}
\end{ruledtabular}
\end{table}

\begin{table}[!htbp]
\caption{\label{tab:cross_sections}
Differential cross sections at 61.5 and 97.1~MeV together with the statistical and total point-to-point systematic uncertainties.} 
\begin{ruledtabular}
\begin{tabular}{cccccc}
 $\theta_{\text{lab}}$ & $\phi$ & $d\sigma/d\Omega$ & Stat & Point-to-point  \\
 (deg) & (deg) & (nb/sr) & (nb/sr) & Syst (nb/sr) \\
\hline
\multicolumn{5}{c}{Circularly polarized gamma-ray beam at 61.5~MeV} \\
\hline
40.3  & 0   & 62.2 & $\pm$ 1.7 & $\pm$ 1.4  \\
55.1  & 180 & 49.8 & $\pm$ 1.7 & $\pm$ 1.1  \\
55.5  & 270 & 51.1 & $\pm$ 1.7 & $\pm$ 0.8  \\
89.6  & 180 & 34.9 & $\pm$ 1.6 & $\pm$ 0.9  \\
90.3  & 270 & 38.5 & $\pm$ 1.7 & $\pm$ 1.5  \\
113.8 & 0   & 41.9 & $\pm$ 0.9 & $\pm$ 1.1  \\
149.3 & 180 & 72.8 & $\pm$ 1.4 & $\pm$ 1.0  \\
\hline
\multicolumn{5}{c}{Circularly polarized gamma-ray beam at 97.1~MeV} \\
\hline
40.0  & 0   & 43.6 & $\pm$ 2.4 & $\pm$ 1.5  \\
55.0  & 180 & 39.6 & $\pm$ 2.8 & $\pm$ 2.3  \\
55.0  & 270 & 36.4 & $\pm$ 2.7 & $\pm$ 1.9  \\
90.0  & 180 & 32.3 & $\pm$ 2.0 & $\pm$ 2.0  \\
90.0  & 270 & 31.1 & $\pm$ 2.6 & $\pm$ 2.6  \\
115.0 & 0   & 38.4 & $\pm$ 1.4 & $\pm$ 1.0  \\
150.0 & 180 & 51.1 & $\pm$ 1.7 & $\pm$ 1.1  \\
\end{tabular}
\end{ruledtabular}
\end{table}

\begin{figure}[!htbp]
\includegraphics[width=\linewidth]{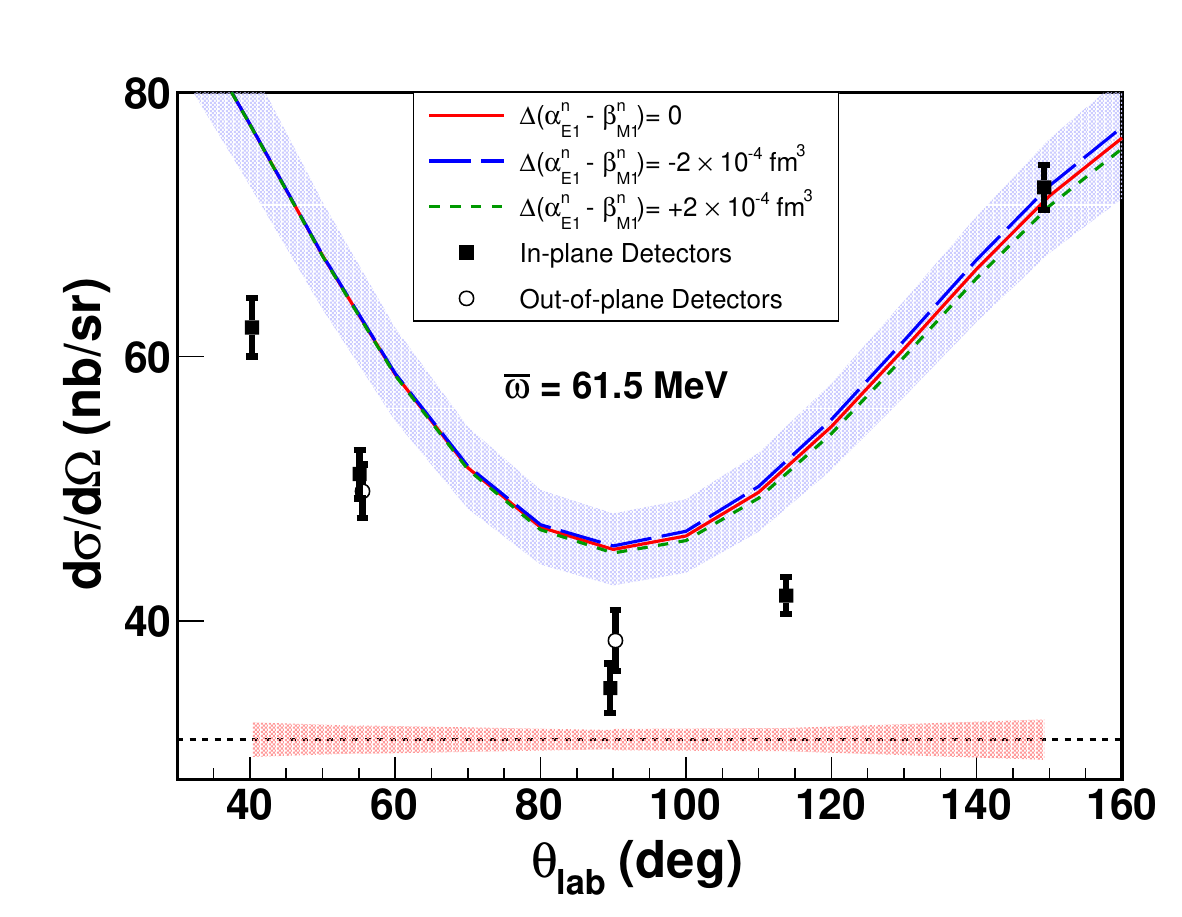}
\includegraphics[width=\linewidth]{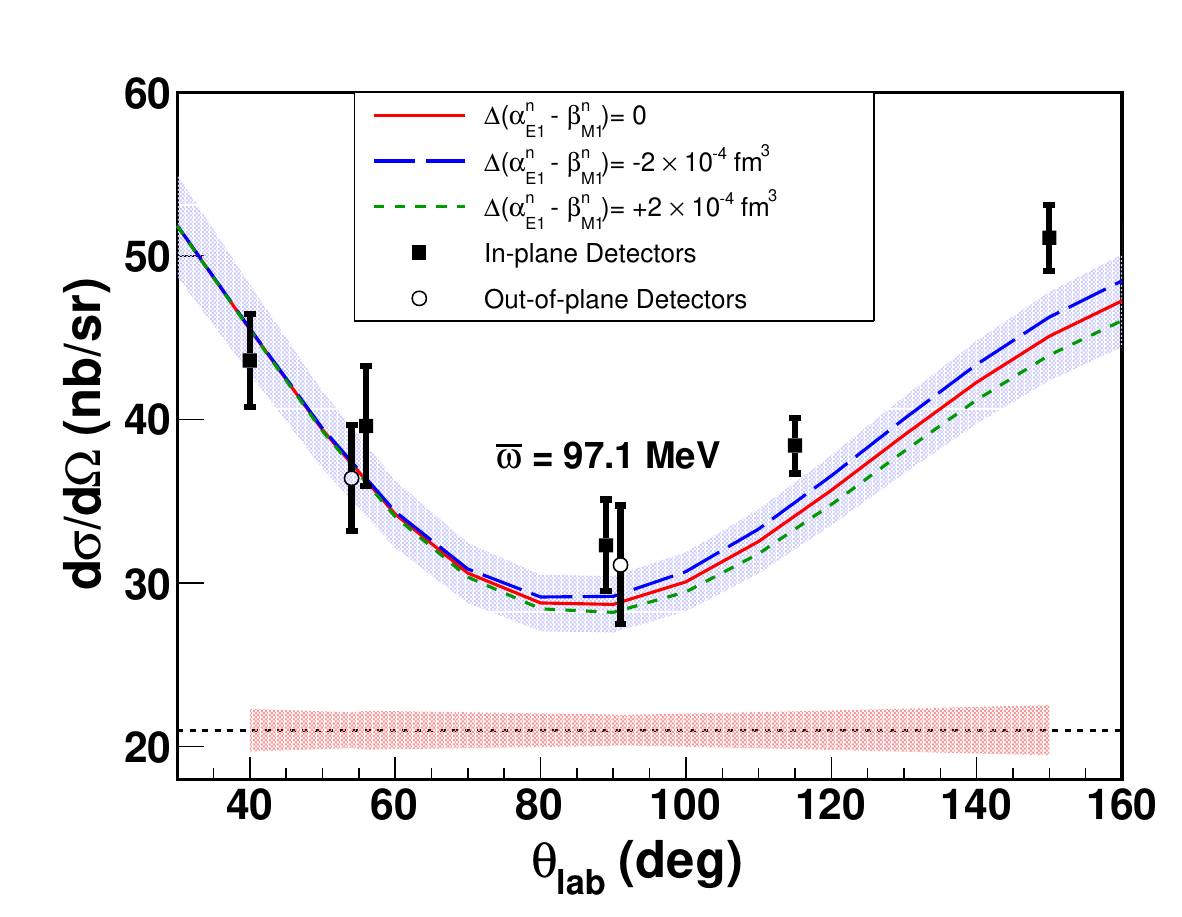}
\caption{\label{fig:dxs_exp} Differential cross sections for elastic Compton scattering from $^3$He at intensity-weighted incident photon beam energies of 61.5 (top) and 97.1 MeV (bottom). Black squares and open circles represent data from the in-plane and out-of-plane detectors, respectively. At 97.1 MeV, open circles at 55 and 90 degrees are slightly shifted for clarity. Error bars indicate the quadratic sum of the statistical and total point-to-point systematic uncertainties. At the bottom of each panel, the light-pink band (\textcolor{purple!40}{\rule{1.5ex}{1.5ex}}) represents the overall normalization systematic uncertainty. The solid red curve represents the $\chi$EFT calculation~\cite{Margaryan:2018opu} when using $\alpha^{n}_{E1}$ and $\beta^{n}_{M1}$ values from Ref.~\cite{COMPTONMAX-lab:2014cve}. The short-dashed green and long-dashed blue curves demonstrate the effect of varying the polarizability difference, $\alpha^{n}_{E1}-\beta^{n}_{M1}$, by $\pm 2 \times 10^{-4}$ fm$^3$. 
The shaded purple band (\textcolor{bandpurple}{\rule{1.5ex}{1.5ex}}) surrounding the red, blue, and green curves represents the $\pm$6\% uncertainty from Ref.~\cite{Margaryan:2018opu}, which is angular-dependent.} 

\end{figure}

The differential cross sections for elastic Compton scattering from $^3$He and the associated uncertainties are presented in Fig.~\ref{fig:dxs_exp} and summarized in Table~\ref{tab:cross_sections}. Theoretical predictions based on $\chi$EFT~\cite{Margaryan:2018opu} are overlaid in Fig.~\ref{fig:dxs_exp} to compare with the experimental results and to illustrate the sensitivity to $\alpha^{n}_{E1}-\beta^{n}_{M1}$. This calculation uses the central values of $\alpha^{n}_{E1}$ and $\beta^{n}_{M1}$ from Ref.~\cite{COMPTONMAX-lab:2014cve} rather than the PDG values in Eq.(\ref{eq:PDG_2025}). However, the difference is not significant here. Being practically insensitive to the polarizability difference, the forward-angle data provide crucial constraints on the overall normalization of the angular distribution. These measurements are therefore essential for benchmarking the theoretical framework prior to extracting $\alpha_{E1}^{n}-\beta_{M1}^{n}$.
Notably, the agreement in magnitude with $\chi$EFT at 97.1 MeV demonstrates that the Compton-scattering cross section at this energy is primarily driven by meson-exchange currents and scales with the number of neutron-proton ($np$) pairs~\cite{Margaryan:2018opu,long2025scattering}. This validates the theoretical treatment of nuclear dynamics in light nuclei and complements evidence for the $np$-pair dominance observed in high-energy electron-scattering experiments~\cite{ye2018search,li2022revealing,Hen:2026hyi}.

On the other hand, most data points at 61.5 MeV fall below the theoretical predictions by about 20\%. These differences are larger than theoretical uncertainties obtained from the order-by-order behavior of the EFT expansion for Compton scattering at photon energies of order the pion mass~\cite{Margaryan:2018opu}. 
However, this discrepancy was expected because the existing $\chi$EFT calculation for $^3$He is organized (i.e. uses a power counting) appropriate for photon energies $\sim$100 MeV~\cite{Margaryan:2018opu}. For energies $\sim$40~MeV, collective rescattering effects need to be added to the amplitude, and that theoretical work is underway~\cite{Walet:2023myk,Walet:inprep}. 
These are expected to reduce the cross section at lower energies, decreasing the magnitude while mostly preserving the angular dependence~\cite{Hildebrandt:2010cqd}. We reiterate that data at forward angles is a check of whether theory predicts the correct cross-section size, but results there are insensitive to $\alpha_{E1}^{n}-\beta_{M1}^{n}$.
The forward data at 61.5~MeV are clearly incompatible with the $\chi$EFT computation without rescattering, but the sensitivity to polarizabilities is small there. On the other hand, forward data and theory match well at 97.1~MeV, where the sensitivity to the polarizabilities is at least twice as large (see Fig.~\ref{fig:dxs_exp}). Therefore, our data will be the foundation of an independent high-accuracy extraction once the theory has been refined.

While accurately extracting polarizabilities from this work will require extending the current $\chi$EFT calculations to one higher order and incorporating mechanisms crucial below 50 MeV, this experiment stands as a major milestone in the study of neutron polarizabilities. It is not only the world’s first measurement of elastic Compton scattering on $^3$He, but it also achieves significantly lower relative uncertainties than the benchmark deuteron measurement by Myers \textit{et al.}~\cite{Myers:2015aba}. In absolute terms, the uncertainties reported here are comparable to that benchmark and strictly lower than all other measurements on the deuteron, $^4$He, and $^6$Li to date. This precision is particularly valuable because the absolute differential cross section of $^3$He exhibits greater sensitivity to $\alpha_{E1}^{n}-\beta_{M1}^{n}$ than that of the deuteron~\cite{choudhury2007investigating,Margaryan:2018opu}. Therefore, once the theory has been improved in the two ways described above, this first-ever $^3$He Compton dataset should enable an independent and precise extraction of this polarizability difference. 
This extracted value will then offer a rigorous point of comparison for the latest, rapidly advancing predictions from lattice QCD~\cite{wang2024nucleon}.
In summary, nucleon polarizabilities are important observables that provide a critical bridge between the low-energy regime of $\chi$EFT and first-principles lattice QCD calculations, making their precise extraction via Compton scattering an essential test of QCD in the non-perturbative regime. In the current experiment, elastic Compton scattering from $^3 \rm He$ has been measured for the first time with high precision. Differential cross sections are reported at laboratory scattering angles of $40^\circ, 55^\circ, 90^\circ, 115^\circ,$ and $150^\circ$ using quasi-monoenergetic, circularly polarized photon beams with intensity-weighted mean energies of 61.5 and 97.1 MeV. 
This first-ever $^3 \rm He$ dataset agrees with key aspects of the $\chi$EFT predictions, validating core ingredients of this approach. The remaining discrepancies motivate the inclusion of rescattering effects in the $\chi$EFT amplitude which restore the Thomson limit, and of higher-order effects more pertinent at high energies.
Furthermore, this experiment demonstrates the HI$\gamma$S facility's potential to conduct high-precision Compton-scattering measurements at higher energies, where cross sections are more sensitive to nucleon polarizabilities. Finally, technical improvements achieved during this work have strongly motivated subsequent experiments at HI$\gamma$S, including elastic Compton scattering from $^4$He at 87 and 100 MeV \cite{lewis2025nucleon}, $^6$Li at $\sim$60~MeV, as well as future measurements with proton, $^3$He, and $^6$Li targets at various energies to enable a comprehensive comparison with theory. 



We acknowledge the support of the HI$\gamma$S accelerator staff for the delivery of high-quality gamma-ray beams and their help with the experimental setup. 
We acknowledge Judith McGovern for insightful discussions on $\chi$EFT calculations for elastic Compton scattering on $^3$He.
We acknowledge Rob Pywell for the support in constructing and testing the new photon beam intensity monitoring system and for providing the simulation package for the system.
This research is supported in part by the U.S. Department of Energy grant numbers DE-FG02-03ER41231, DE-SC0016581, DE-SC0005367, DE-FG02-97ER41033, DE-SC0016656, DE-SC0025227, DE-SC0015393, DE-FG02-93ER40756, the National Science Foundation grant no. 2232117, and  NSERC SAPIN-2022-00025.

\bibliography{main.bib}

\begin{thebibliography}{49}%
\makeatletter
\providecommand \@ifxundefined [1]{%
 \@ifx{#1\undefined}
}%
\providecommand \@ifnum [1]{%
 \ifnum #1\expandafter \@firstoftwo
 \else \expandafter \@secondoftwo
 \fi
}%
\providecommand \@ifx [1]{%
 \ifx #1\expandafter \@firstoftwo
 \else \expandafter \@secondoftwo
 \fi
}%
\providecommand \natexlab [1]{#1}%
\providecommand \enquote  [1]{``#1''}%
\providecommand \bibnamefont  [1]{#1}%
\providecommand \bibfnamefont [1]{#1}%
\providecommand \citenamefont [1]{#1}%
\providecommand \href@noop [0]{\@secondoftwo}%
\providecommand \href [0]{\begingroup \@sanitize@url \@href}%
\providecommand \@href[1]{\@@startlink{#1}\@@href}%
\providecommand \@@href[1]{\endgroup#1\@@endlink}%
\providecommand \@sanitize@url [0]{\catcode `\\12\catcode `\$12\catcode `\&12\catcode `\#12\catcode `\^12\catcode `\_12\catcode `\%12\relax}%
\providecommand \@@startlink[1]{}%
\providecommand \@@endlink[0]{}%
\providecommand \url  [0]{\begingroup\@sanitize@url \@url }%
\providecommand \@url [1]{\endgroup\@href {#1}{\urlprefix }}%
\providecommand \urlprefix  [0]{URL }%
\providecommand \Eprint [0]{\href }%
\providecommand \doibase [0]{https://doi.org/}%
\providecommand \selectlanguage [0]{\@gobble}%
\providecommand \bibinfo  [0]{\@secondoftwo}%
\providecommand \bibfield  [0]{\@secondoftwo}%
\providecommand \translation [1]{[#1]}%
\providecommand \BibitemOpen [0]{}%
\providecommand \bibitemStop [0]{}%
\providecommand \bibitemNoStop [0]{.\EOS\space}%
\providecommand \EOS [0]{\spacefactor3000\relax}%
\providecommand \BibitemShut  [1]{\csname bibitem#1\endcsname}%
\let\auto@bib@innerbib\@empty
\bibitem [{\citenamefont {Schumacher}(2019)}]{Schumacher:2019ikn}%
  \BibitemOpen
  \bibfield  {author} {\bibinfo {author} {\bibfnamefont {M.}~\bibnamefont {Schumacher}},\ }\bibfield  {title} {\bibinfo {title} {{Polarizability of the nucleon}},\ }\href {https://doi.org/10.31526/lhep.4.2019.136} {\bibfield  {journal} {\bibinfo  {journal} {LHEP}\ }\textbf {\bibinfo {volume} {2}},\ \bibinfo {pages} {4} (\bibinfo {year} {2019})}\BibitemShut {NoStop}%
\bibitem [{\citenamefont {Griesshammer}\ \emph {et~al.}(2012)\citenamefont {Griesshammer} \emph {et~al.}}]{Griesshammer12}%
  \BibitemOpen
  \bibfield  {author} {\bibinfo {author} {\bibfnamefont {H.~W.}\ \bibnamefont {Griesshammer}} \emph {et~al.},\ }\bibfield  {title} {\bibinfo {title} {{Using effective field theory to analyse low-energy Compton scattering data from protons and light nuclei}},\ }\href {https://doi.org/10.1016/j.ppnp.2012.04.003} {\bibfield  {journal} {\bibinfo  {journal} {Prog. Part. Nucl. Phys.}\ }\textbf {\bibinfo {volume} {67}},\ \bibinfo {pages} {841} (\bibinfo {year} {2012})}\BibitemShut {NoStop}%
\bibitem [{\citenamefont {Detmold}\ \emph {et~al.}(2006)\citenamefont {Detmold}, \citenamefont {Tiburzi},\ and\ \citenamefont {Walker-Loud}}]{detmold2006electromagnetic}%
  \BibitemOpen
  \bibfield  {author} {\bibinfo {author} {\bibfnamefont {W.}~\bibnamefont {Detmold}}, \bibinfo {author} {\bibfnamefont {B.}~\bibnamefont {Tiburzi}},\ and\ \bibinfo {author} {\bibfnamefont {A.}~\bibnamefont {Walker-Loud}},\ }\bibfield  {title} {\bibinfo {title} {Electromagnetic and spin polarizabilities in {lattice QCD}},\ }\href@noop {} {\bibfield  {journal} {\bibinfo  {journal} {Phys. Rev. D}\ }\textbf {\bibinfo {volume} {73}},\ \bibinfo {pages} {114505} (\bibinfo {year} {2006})}\BibitemShut {NoStop}%
\bibitem [{\citenamefont {Detmold}\ \emph {et~al.}(2010)\citenamefont {Detmold}, \citenamefont {Tiburzi},\ and\ \citenamefont {Walker-Loud}}]{detmold2010extracting}%
  \BibitemOpen
  \bibfield  {author} {\bibinfo {author} {\bibfnamefont {W.}~\bibnamefont {Detmold}}, \bibinfo {author} {\bibfnamefont {B.~C.}\ \bibnamefont {Tiburzi}},\ and\ \bibinfo {author} {\bibfnamefont {A.}~\bibnamefont {Walker-Loud}},\ }\bibfield  {title} {\bibinfo {title} {{Extracting Nucleon Magnetic Moments and Electric Polarizabilities from Lattice QCD in Background Electric Fields}},\ }\href {https://doi.org/10.1103/PhysRevD.81.054502} {\bibfield  {journal} {\bibinfo  {journal} {Phys. Rev. D}\ }\textbf {\bibinfo {volume} {81}},\ \bibinfo {pages} {054502} (\bibinfo {year} {2010})},\ \Eprint {https://arxiv.org/abs/1001.1131} {arXiv:1001.1131 [hep-lat]} \BibitemShut {NoStop}%
\bibitem [{\citenamefont {Bignell}\ \emph {et~al.}(2020)\citenamefont {Bignell}, \citenamefont {Kamleh},\ and\ \citenamefont {Leinweber}}]{Bignell:2020xkf}%
  \BibitemOpen
  \bibfield  {author} {\bibinfo {author} {\bibfnamefont {R.}~\bibnamefont {Bignell}}, \bibinfo {author} {\bibfnamefont {W.}~\bibnamefont {Kamleh}},\ and\ \bibinfo {author} {\bibfnamefont {D.}~\bibnamefont {Leinweber}},\ }\bibfield  {title} {\bibinfo {title} {{Magnetic polarizability of the nucleon using a Laplacian mode projection}},\ }\href {https://doi.org/10.1103/PhysRevD.101.094502} {\bibfield  {journal} {\bibinfo  {journal} {Phys. Rev. D}\ }\textbf {\bibinfo {volume} {101}},\ \bibinfo {pages} {094502} (\bibinfo {year} {2020})}\BibitemShut {NoStop}%
\bibitem [{\citenamefont {Hall}\ \emph {et~al.}(2014)\citenamefont {Hall}, \citenamefont {Leinweber},\ and\ \citenamefont {Young}}]{Hall:2013dva}%
  \BibitemOpen
  \bibfield  {author} {\bibinfo {author} {\bibfnamefont {J.~M.~M.}\ \bibnamefont {Hall}}, \bibinfo {author} {\bibfnamefont {D.~B.}\ \bibnamefont {Leinweber}},\ and\ \bibinfo {author} {\bibfnamefont {R.~D.}\ \bibnamefont {Young}},\ }\bibfield  {title} {\bibinfo {title} {{Finite-volume and partial quenching effects in the magnetic polarizability of the neutron}},\ }\href {https://doi.org/10.1103/PhysRevD.89.054511} {\bibfield  {journal} {\bibinfo  {journal} {Phys. Rev. D}\ }\textbf {\bibinfo {volume} {89}},\ \bibinfo {pages} {054511} (\bibinfo {year} {2014})}\BibitemShut {NoStop}%
\bibitem [{\citenamefont {Lujan}\ \emph {et~al.}(2016)\citenamefont {Lujan} \emph {et~al.}}]{Lujan:2016ffj}%
  \BibitemOpen
  \bibfield  {author} {\bibinfo {author} {\bibfnamefont {M.}~\bibnamefont {Lujan}} \emph {et~al.},\ }\bibfield  {title} {\bibinfo {title} {{Finite volume effects on the electric polarizability of neutral hadrons in lattice QCD}},\ }\href {https://doi.org/10.1103/PhysRevD.94.074506} {\bibfield  {journal} {\bibinfo  {journal} {Phys. Rev. D}\ }\textbf {\bibinfo {volume} {94}},\ \bibinfo {pages} {074506} (\bibinfo {year} {2016})}\BibitemShut {NoStop}%
\bibitem [{\citenamefont {Engelhardt}(2011)}]{Engelhardt:2011qq}%
  \BibitemOpen
  \bibfield  {author} {\bibinfo {author} {\bibfnamefont {M.}~\bibnamefont {Engelhardt}},\ }\bibfield  {title} {\bibinfo {title} {{Exploration of the electric spin polarizability of the neutron in lattice {QCD}}},\ }\href {https://doi.org/10.22323/1.139.0153} {\bibfield  {journal} {\bibinfo  {journal} {PoS}\ }\textbf {\bibinfo {volume} {LATTICE2011}},\ \bibinfo {pages} {153} (\bibinfo {year} {2011})}\BibitemShut {NoStop}%
\bibitem [{\citenamefont {Saenz}\ \emph {et~al.}(2021)\citenamefont {Saenz}, \citenamefont {Engelhardt},\ and\ \citenamefont {H{\"o}llwieser}}]{Saenz:2020yxy}%
  \BibitemOpen
  \bibfield  {author} {\bibinfo {author} {\bibfnamefont {J.}~\bibnamefont {Saenz}}, \bibinfo {author} {\bibfnamefont {M.}~\bibnamefont {Engelhardt}},\ and\ \bibinfo {author} {\bibfnamefont {R.}~\bibnamefont {H{\"o}llwieser}},\ }\bibfield  {title} {\bibinfo {title} {{Energy of a pointlike neutron in an external electromagnetic field}},\ }\href {https://doi.org/10.1103/PhysRevD.104.056002} {\bibfield  {journal} {\bibinfo  {journal} {Phys. Rev. D}\ }\textbf {\bibinfo {volume} {104}},\ \bibinfo {pages} {056002} (\bibinfo {year} {2021})}\BibitemShut {NoStop}%
\bibitem [{\citenamefont {Lee}\ \emph {et~al.}(2024)\citenamefont {Lee} \emph {et~al.}}]{Lee:2024dgf}%
  \BibitemOpen
  \bibfield  {author} {\bibinfo {author} {\bibfnamefont {F.~X.}\ \bibnamefont {Lee}} \emph {et~al.},\ }\bibfield  {title} {\bibinfo {title} {{Neutral pion polarizabilities from four-point functions in lattice {QCD}}},\ }\href {https://doi.org/10.1016/j.nuclphysb.2024.116701} {\bibfield  {journal} {\bibinfo  {journal} {Nucl. Phys. B}\ }\textbf {\bibinfo {volume} {1008}},\ \bibinfo {pages} {116701} (\bibinfo {year} {2024})}\BibitemShut {NoStop}%
\bibitem [{\citenamefont {Wang}\ \emph {et~al.}(2024)\citenamefont {Wang} \emph {et~al.}}]{wang2024nucleon}%
  \BibitemOpen
  \bibfield  {author} {\bibinfo {author} {\bibfnamefont {X.-H.}\ \bibnamefont {Wang}} \emph {et~al.},\ }\bibfield  {title} {\bibinfo {title} {Nucleon electric polarizabilities and nucleon-pion scattering at the physical pion mass},\ }\href@noop {} {\bibfield  {journal} {\bibinfo  {journal} {Phys. Rev. Lett.}\ }\textbf {\bibinfo {volume} {133}},\ \bibinfo {pages} {141901} (\bibinfo {year} {2024})}\BibitemShut {NoStop}%
\bibitem [{\citenamefont {Drechsel}\ \emph {et~al.}(2003)\citenamefont {Drechsel}, \citenamefont {Pasquini},\ and\ \citenamefont {Vanderhaeghen}}]{drechsel2003dispersion}%
  \BibitemOpen
  \bibfield  {author} {\bibinfo {author} {\bibfnamefont {D.}~\bibnamefont {Drechsel}}, \bibinfo {author} {\bibfnamefont {B.}~\bibnamefont {Pasquini}},\ and\ \bibinfo {author} {\bibfnamefont {M.}~\bibnamefont {Vanderhaeghen}},\ }\bibfield  {title} {\bibinfo {title} {Dispersion relations in real and virtual {Compton} scattering},\ }\href@noop {} {\bibfield  {journal} {\bibinfo  {journal} {Physics Reports}\ }\textbf {\bibinfo {volume} {378}},\ \bibinfo {pages} {99} (\bibinfo {year} {2003})}\BibitemShut {NoStop}%
\bibitem [{\citenamefont {Hagelstein}\ \emph {et~al.}(2016)\citenamefont {Hagelstein}, \citenamefont {Miskimen},\ and\ \citenamefont {Pascalutsa}}]{Hagelstein15}%
  \BibitemOpen
  \bibfield  {author} {\bibinfo {author} {\bibfnamefont {F.}~\bibnamefont {Hagelstein}}, \bibinfo {author} {\bibfnamefont {R.}~\bibnamefont {Miskimen}},\ and\ \bibinfo {author} {\bibfnamefont {V.}~\bibnamefont {Pascalutsa}},\ }\bibfield  {title} {\bibinfo {title} {{Nucleon Polarizabilities: from Compton Scattering to Hydrogen Atom}},\ }\href {https://doi.org/10.1016/j.ppnp.2015.12.001} {\bibfield  {journal} {\bibinfo  {journal} {Prog. Part. Nucl. Phys.}\ }\textbf {\bibinfo {volume} {88}},\ \bibinfo {pages} {29} (\bibinfo {year} {2016})}\BibitemShut {NoStop}%
\bibitem [{\citenamefont {Hagelstein}(2020)}]{Hagelstein:2020vog}%
  \BibitemOpen
  \bibfield  {author} {\bibinfo {author} {\bibfnamefont {F.}~\bibnamefont {Hagelstein}},\ }\bibfield  {title} {\bibinfo {title} {{Nucleon Polarizabilities and Compton Scattering as Playground for Chiral Perturbation Theory}},\ }\href {https://doi.org/10.3390/sym12091407} {\bibfield  {journal} {\bibinfo  {journal} {Symmetry}\ }\textbf {\bibinfo {volume} {12}},\ \bibinfo {pages} {1407} (\bibinfo {year} {2020})}\BibitemShut {NoStop}%
\bibitem [{\citenamefont {Maximon}(1989)}]{maximon1989scattering}%
  \BibitemOpen
  \bibfield  {author} {\bibinfo {author} {\bibfnamefont {L.~C.}\ \bibnamefont {Maximon}},\ }\bibfield  {title} {\bibinfo {title} {Scattering of polarized photons by protons},\ }\href@noop {} {\bibfield  {journal} {\bibinfo  {journal} {Phys. Rev. C}\ }\textbf {\bibinfo {volume} {39}},\ \bibinfo {pages} {347} (\bibinfo {year} {1989})}\BibitemShut {NoStop}%
\bibitem [{\citenamefont {Baldin}(1960)}]{baldin1960polarizability}%
  \BibitemOpen
  \bibfield  {author} {\bibinfo {author} {\bibfnamefont {A.}~\bibnamefont {Baldin}},\ }\bibfield  {title} {\bibinfo {title} {Polarizability of nucleons},\ }\href@noop {} {\bibfield  {journal} {\bibinfo  {journal} {Nuclear Physics}\ }\textbf {\bibinfo {volume} {18}},\ \bibinfo {pages} {310} (\bibinfo {year} {1960})}\BibitemShut {NoStop}%
\bibitem [{\citenamefont {Navas}\ \emph {et~al.}(2024)\citenamefont {Navas} \emph {et~al.}}]{ParticleDataGroup:2024cfk}%
  \BibitemOpen
  \bibfield  {author} {\bibinfo {author} {\bibfnamefont {S.}~\bibnamefont {Navas}} \emph {et~al.} (\bibinfo {collaboration} {Particle Data Group}),\ }\bibfield  {title} {\bibinfo {title} {{Review of particle physics}},\ }\href {https://doi.org/10.1103/PhysRevD.110.030001} {\bibfield  {journal} {\bibinfo  {journal} {Phys. Rev. D}\ }\textbf {\bibinfo {volume} {110}},\ \bibinfo {pages} {030001} (\bibinfo {year} {2024})}\BibitemShut {NoStop}%
\bibitem [{\citenamefont {Gasser}\ \emph {et~al.}(2015)\citenamefont {Gasser} \emph {et~al.}}]{Gasser:2015dwa}%
  \BibitemOpen
  \bibfield  {author} {\bibinfo {author} {\bibfnamefont {J.}~\bibnamefont {Gasser}} \emph {et~al.},\ }\bibfield  {title} {\bibinfo {title} {{Cottingham formula and nucleon polarisabilities}},\ }\href {https://doi.org/10.1140/epjc/s10052-015-3580-9} {\bibfield  {journal} {\bibinfo  {journal} {Eur. Phys. J. C}\ }\textbf {\bibinfo {volume} {75}},\ \bibinfo {pages} {375} (\bibinfo {year} {2015})},\ \bibinfo {note} {[Erratum: Eur.Phys.J.C 80, 353 (2020)]}\BibitemShut {NoStop}%
\bibitem [{\citenamefont {Long}\ and\ \citenamefont {Griesshammer}(2025)}]{long2025scattering}%
  \BibitemOpen
  \bibfield  {author} {\bibinfo {author} {\bibfnamefont {A.}~\bibnamefont {Long}}\ and\ \bibinfo {author} {\bibfnamefont {H.~W.}\ \bibnamefont {Griesshammer}},\ }\bibfield  {title} {\bibinfo {title} {Scattering observables from few-body densities and {Compton} scattering on {$^6$Li}},\ }\href@noop {} {\bibfield  {journal} {\bibinfo  {journal} {arXiv:2504.00989}\ } (\bibinfo {year} {2025})}\BibitemShut {NoStop}%
\bibitem [{\citenamefont {Lucas}(1994)}]{lucas1994compton}%
  \BibitemOpen
  \bibfield  {author} {\bibinfo {author} {\bibfnamefont {M.~A.}\ \bibnamefont {Lucas}},\ }\emph {\bibinfo {title} {{Compton scattering from the deuteron at intermediate energies}}},\ \href@noop {} {Ph.D. thesis},\ \bibinfo  {school} {University of Illinois at Urbana-Champaign} (\bibinfo {year} {1994})\BibitemShut {NoStop}%
\bibitem [{\citenamefont {Hornidge}\ \emph {et~al.}(2000)\citenamefont {Hornidge} \emph {et~al.}}]{Hornidge:1999xs}%
  \BibitemOpen
  \bibfield  {author} {\bibinfo {author} {\bibfnamefont {D.~L.}\ \bibnamefont {Hornidge}} \emph {et~al.},\ }\bibfield  {title} {\bibinfo {title} {{Elastic Compton scattering from the deuteron and nucleon polarizabilities}},\ }\href {https://doi.org/10.1103/PhysRevLett.84.2334} {\bibfield  {journal} {\bibinfo  {journal} {Phys. Rev. Lett.}\ }\textbf {\bibinfo {volume} {84}},\ \bibinfo {pages} {2334} (\bibinfo {year} {2000})}\BibitemShut {NoStop}%
\bibitem [{\citenamefont {Lundin}\ \emph {et~al.}(2003)\citenamefont {Lundin} \emph {et~al.}}]{Lundin:2002jy}%
  \BibitemOpen
  \bibfield  {author} {\bibinfo {author} {\bibfnamefont {M.}~\bibnamefont {Lundin}} \emph {et~al.},\ }\bibfield  {title} {\bibinfo {title} {{Compton scattering from the deuteron and neutron polarizabilities}},\ }\href {https://doi.org/10.1103/PhysRevLett.90.192501} {\bibfield  {journal} {\bibinfo  {journal} {Phys. Rev. Lett.}\ }\textbf {\bibinfo {volume} {90}},\ \bibinfo {pages} {192501} (\bibinfo {year} {2003})}\BibitemShut {NoStop}%
\bibitem [{\citenamefont {Myers}\ \emph {et~al.}(2015)\citenamefont {Myers} \emph {et~al.}}]{Myers:2015aba}%
  \BibitemOpen
  \bibfield  {author} {\bibinfo {author} {\bibfnamefont {L.~S.}\ \bibnamefont {Myers}} \emph {et~al.},\ }\bibfield  {title} {\bibinfo {title} {{Compton Scattering from the Deuteron below Pion-Production Threshold}},\ }\href {https://doi.org/10.1103/PhysRevC.92.025203} {\bibfield  {journal} {\bibinfo  {journal} {Phys. Rev. C}\ }\textbf {\bibinfo {volume} {92}},\ \bibinfo {pages} {025203} (\bibinfo {year} {2015})}\BibitemShut {NoStop}%
\bibitem [{\citenamefont {Myers}\ \emph {et~al.}(2014{\natexlab{a}})\citenamefont {Myers} \emph {et~al.}}]{COMPTONMAX-lab:2014cve}%
  \BibitemOpen
  \bibfield  {author} {\bibinfo {author} {\bibfnamefont {L.~S.}\ \bibnamefont {Myers}} \emph {et~al.} (\bibinfo {collaboration} {COMPTON@MAX-lab}),\ }\bibfield  {title} {\bibinfo {title} {{Measurement of Compton Scattering from the Deuteron and an Improved Extraction of the Neutron Electromagnetic Polarizabilities}},\ }\href {https://doi.org/10.1103/PhysRevLett.113.262506} {\bibfield  {journal} {\bibinfo  {journal} {Phys. Rev. Lett.}\ }\textbf {\bibinfo {volume} {113}},\ \bibinfo {pages} {262506} (\bibinfo {year} {2014}{\natexlab{a}})}\BibitemShut {NoStop}%
\bibitem [{\citenamefont {Weller}\ \emph {et~al.}(2009)\citenamefont {Weller} \emph {et~al.}}]{Weller:2009zza}%
  \BibitemOpen
  \bibfield  {author} {\bibinfo {author} {\bibfnamefont {H.~R.}\ \bibnamefont {Weller}} \emph {et~al.},\ }\bibfield  {title} {\bibinfo {title} {{Research opportunities at the upgraded {HI$\gamma$S} facility}},\ }\href {https://doi.org/10.1016/j.ppnp.2008.07.001} {\bibfield  {journal} {\bibinfo  {journal} {Prog. Part. Nucl. Phys.}\ }\textbf {\bibinfo {volume} {62}},\ \bibinfo {pages} {257} (\bibinfo {year} {2009})}\BibitemShut {NoStop}%
\bibitem [{\citenamefont {Li}\ \emph {et~al.}(2022{\natexlab{a}})\citenamefont {Li} \emph {et~al.}}]{Li:2022vnz}%
  \BibitemOpen
  \bibfield  {author} {\bibinfo {author} {\bibfnamefont {X.}~\bibnamefont {Li}} \emph {et~al.},\ }\bibfield  {title} {\bibinfo {title} {{Proton Compton Scattering from Linearly Polarized Gamma Rays}},\ }\href {https://doi.org/10.1103/PhysRevLett.128.132502} {\bibfield  {journal} {\bibinfo  {journal} {Phys. Rev. Lett.}\ }\textbf {\bibinfo {volume} {128}},\ \bibinfo {pages} {132502} (\bibinfo {year} {2022}{\natexlab{a}})}\BibitemShut {NoStop}%
\bibitem [{\citenamefont {Godagama}(2022)}]{Godagama:2022aiy}%
  \BibitemOpen
  \bibfield  {author} {\bibinfo {author} {\bibfnamefont {D.}~\bibnamefont {Godagama}},\ }\emph {\bibinfo {title} {{Elastic and Inelastic Compton Scattering from Deuterium at 61 MeV}}},\ \href {https://doi.org/10.13023/etd.2022.281} {Ph.D. thesis},\ \bibinfo  {school} {University of Kentucky} (\bibinfo {year} {2022})\BibitemShut {NoStop}%
\bibitem [{\citenamefont {Li}\ \emph {et~al.}(2020)\citenamefont {Li} \emph {et~al.}}]{Li:2019irp}%
  \BibitemOpen
  \bibfield  {author} {\bibinfo {author} {\bibfnamefont {X.}~\bibnamefont {Li}} \emph {et~al.},\ }\bibfield  {title} {\bibinfo {title} {{Compton scattering from $^4$He at the TUNL HI$\gamma$S facility}},\ }\href {https://doi.org/10.1103/PhysRevC.101.034618} {\bibfield  {journal} {\bibinfo  {journal} {Phys. Rev. C}\ }\textbf {\bibinfo {volume} {101}},\ \bibinfo {pages} {034618} (\bibinfo {year} {2020})}\BibitemShut {NoStop}%
\bibitem [{\citenamefont {Myers}\ \emph {et~al.}(2012)\citenamefont {Myers} \emph {et~al.}}]{Myers:2012xw}%
  \BibitemOpen
  \bibfield  {author} {\bibinfo {author} {\bibfnamefont {L.~S.}\ \bibnamefont {Myers}} \emph {et~al.},\ }\bibfield  {title} {\bibinfo {title} {{Compton scattering from {$^6$Li} at 60 MeV}},\ }\href {https://doi.org/10.1103/PhysRevC.86.044614} {\bibfield  {journal} {\bibinfo  {journal} {Phys. Rev. C}\ }\textbf {\bibinfo {volume} {86}},\ \bibinfo {pages} {044614} (\bibinfo {year} {2012})}\BibitemShut {NoStop}%
\bibitem [{\citenamefont {Myers}\ \emph {et~al.}(2014{\natexlab{b}})\citenamefont {Myers} \emph {et~al.}}]{myers2014compton}%
  \BibitemOpen
  \bibfield  {author} {\bibinfo {author} {\bibfnamefont {L.}~\bibnamefont {Myers}} \emph {et~al.},\ }\bibfield  {title} {\bibinfo {title} {Compton scattering from {$^6$Li} at 86 {MeV}},\ }\href@noop {} {\bibfield  {journal} {\bibinfo  {journal} {Phys. Rev. C}\ }\textbf {\bibinfo {volume} {90}},\ \bibinfo {pages} {027603} (\bibinfo {year} {2014}{\natexlab{b}})}\BibitemShut {NoStop}%
\bibitem [{\citenamefont {Choudhury}\ \emph {et~al.}(2007)\citenamefont {Choudhury}, \citenamefont {Nogga},\ and\ \citenamefont {Phillips}}]{choudhury2007investigating}%
  \BibitemOpen
  \bibfield  {author} {\bibinfo {author} {\bibfnamefont {D.}~\bibnamefont {Choudhury}}, \bibinfo {author} {\bibfnamefont {A.}~\bibnamefont {Nogga}},\ and\ \bibinfo {author} {\bibfnamefont {D.~R.}\ \bibnamefont {Phillips}},\ }\bibfield  {title} {\bibinfo {title} {Investigating neutron polarizabilities through compton scattering on he 3},\ }\href@noop {} {\bibfield  {journal} {\bibinfo  {journal} {Phys. Rev. Lett.}\ }\textbf {\bibinfo {volume} {98}},\ \bibinfo {pages} {232303} (\bibinfo {year} {2007})}\BibitemShut {NoStop}%
\bibitem [{\citenamefont {Margaryan}\ \emph {et~al.}(2018)\citenamefont {Margaryan} \emph {et~al.}}]{Margaryan:2018opu}%
  \BibitemOpen
  \bibfield  {author} {\bibinfo {author} {\bibfnamefont {A.}~\bibnamefont {Margaryan}} \emph {et~al.},\ }\bibfield  {title} {\bibinfo {title} {{Elastic Compton scattering from$^{3}$He and the role of the Delta}},\ }\href {https://doi.org/10.1140/epja/i2018-12554-x} {\bibfield  {journal} {\bibinfo  {journal} {Eur. Phys. J. A}\ }\textbf {\bibinfo {volume} {54}},\ \bibinfo {pages} {125} (\bibinfo {year} {2018})}\BibitemShut {NoStop}%
\bibitem [{\citenamefont {Mancil}(2026)}]{em2026}%
  \BibitemOpen
  \bibfield  {author} {\bibinfo {author} {\bibfnamefont {E.}~\bibnamefont {Mancil}},\ }\emph {\bibinfo {title} {{Compton Scattering from Liquid \texorpdfstring{\textsuperscript{3}}{3}He with 61.5~MeV Photons}}},\ \href@noop {} {Ph.D. thesis},\ \bibinfo  {school} {Duke U. (main)} (\bibinfo {year} {2026})\BibitemShut {NoStop}%
\bibitem [{\citenamefont {Zhou}(2026)}]{jz2026}%
  \BibitemOpen
  \bibfield  {author} {\bibinfo {author} {\bibfnamefont {J.}~\bibnamefont {Zhou}},\ }\emph {\bibinfo {title} {{Compton Scattering on $^3$He at HI$\gamma$S and Neutron Polarizabilities}}},\ \href@noop {} {Ph.D. thesis},\ \bibinfo  {school} {Duke U. (main)} (\bibinfo {year} {2026})\BibitemShut {NoStop}%
\bibitem [{\citenamefont {Friesen}\ \emph {et~al.}(2026)\citenamefont {Friesen} \emph {et~al.}}]{Friesen_HIgS_Flux}%
  \BibitemOpen
  \bibfield  {author} {\bibinfo {author} {\bibfnamefont {F.}~\bibnamefont {Friesen}} \emph {et~al.},\ }\bibfield  {title} {\bibinfo {title} {A new system for absolute flux measurement at the {High Intensity Gamma-Ray Source}}} (\bibinfo {year} {2026}),\ \bibinfo {note} {to be published}\BibitemShut {NoStop}%
\bibitem [{\citenamefont {Kendellen}\ \emph {et~al.}(2016)\citenamefont {Kendellen} \emph {et~al.}}]{kendellen2016cryogenic}%
  \BibitemOpen
  \bibfield  {author} {\bibinfo {author} {\bibfnamefont {D.}~\bibnamefont {Kendellen}} \emph {et~al.},\ }\bibfield  {title} {\bibinfo {title} {A cryogenic target for {Compton} scattering experiments at {HI$\gamma$S}},\ }\href@noop {} {\bibfield  {journal} {\bibinfo  {journal} {Nucl. Instrum. Meth. A}\ }\textbf {\bibinfo {volume} {840}},\ \bibinfo {pages} {174} (\bibinfo {year} {2016})}\BibitemShut {NoStop}%
\bibitem [{\citenamefont {Leung}(2024)}]{Leung:DNP2024}%
  \BibitemOpen
  \bibfield  {author} {\bibinfo {author} {\bibfnamefont {K.}~\bibnamefont {Leung}},\ }\bibfield  {title} {\bibinfo {title} {Compton@{HIGS}: neutron electromagnetic polarizabilities with helium-3 nuclear {Compton} scattering},\ }in\ \href {https://meetings.aps.org/Meeting/DNP24/Session/P12.7} {\emph {\bibinfo {booktitle} {Bulletin of the American Physical Society}}},\ \bibinfo {series} {2024 Fall Meeting of the APS Division of Nuclear Physics}, Vol.~\bibinfo {volume} {69}\ (\bibinfo {address} {Boston, Massachusetts},\ \bibinfo {year} {2024})\ p.\ \bibinfo {pages} {P12.00007}\BibitemShut {NoStop}%
\bibitem [{\citenamefont {Leung}\ \emph {et~al.}(2026)\citenamefont {Leung} \emph {et~al.}}]{3He_Target}%
  \BibitemOpen
  \bibfield  {author} {\bibinfo {author} {\bibfnamefont {K.~K.~H.}\ \bibnamefont {Leung}} \emph {et~al.},\ }\bibfield  {title} {\bibinfo {title} {A new 0.3~{L} cryogenic liquid {$^3$He} target at {HI$\gamma$S}}} (\bibinfo {year} {2026}),\ \bibinfo {note} {(to be published)}\BibitemShut {NoStop}%
\bibitem [{\citenamefont {Miller}\ \emph {et~al.}(1988)\citenamefont {Miller} \emph {et~al.}}]{Miller:1988gh}%
  \BibitemOpen
  \bibfield  {author} {\bibinfo {author} {\bibfnamefont {J.~P.}\ \bibnamefont {Miller}} \emph {et~al.},\ }\bibfield  {title} {\bibinfo {title} {{{Development of a NaI(Tl) Detector With Superior Photon Energy Resolution for Use Above 200-{MeV}}}},\ }\href {https://doi.org/10.1016/0168-9002(88)90711-5} {\bibfield  {journal} {\bibinfo  {journal} {Nucl. Instrum. Meth. A}\ }\textbf {\bibinfo {volume} {270}},\ \bibinfo {pages} {431} (\bibinfo {year} {1988})}\BibitemShut {NoStop}%
\bibitem [{\citenamefont {Shoniyozov}(2016)}]{shoniyozov2016elastic}%
  \BibitemOpen
  \bibfield  {author} {\bibinfo {author} {\bibfnamefont {K.}~\bibnamefont {Shoniyozov}},\ }\emph {\bibinfo {title} {{Elastic Compton scattering from deuterium near 100 MeV}}},\ \href@noop {} {Ph.D. thesis},\ \bibinfo  {school} {University of Kentucky} (\bibinfo {year} {2016})\BibitemShut {NoStop}%
\bibitem [{\citenamefont {Henshaw}(2010)}]{henshaw2010investigation}%
  \BibitemOpen
  \bibfield  {author} {\bibinfo {author} {\bibfnamefont {S.~S.}\ \bibnamefont {Henshaw}},\ }\emph {\bibinfo {title} {An Investigation of the Isovector Giant Quadrupole Resonance in Bismuth-209 using Polarized Compton Scattering}},\ \href@noop {} {Ph.D. thesis},\ \bibinfo  {school} {Duke University} (\bibinfo {year} {2010})\BibitemShut {NoStop}%
\bibitem [{\citenamefont {Agostinelli}\ \emph {et~al.}(2003)\citenamefont {Agostinelli} \emph {et~al.}}]{GEANT4:2002zbu}%
  \BibitemOpen
  \bibfield  {author} {\bibinfo {author} {\bibfnamefont {S.}~\bibnamefont {Agostinelli}} \emph {et~al.} (\bibinfo {collaboration} {GEANT4}),\ }\bibfield  {title} {\bibinfo {title} {{GEANT4 - A Simulation Toolkit}},\ }\href {https://doi.org/10.1016/S0168-9002(03)01368-8} {\bibfield  {journal} {\bibinfo  {journal} {Nucl. Instrum. Meth. A}\ }\textbf {\bibinfo {volume} {506}},\ \bibinfo {pages} {250} (\bibinfo {year} {2003})}\BibitemShut {NoStop}%
\bibitem [{\citenamefont {Ye}\ \emph {et~al.}(2018)\citenamefont {Ye} \emph {et~al.}}]{ye2018search}%
  \BibitemOpen
  \bibfield  {author} {\bibinfo {author} {\bibfnamefont {Z.}~\bibnamefont {Ye}} \emph {et~al.},\ }\bibfield  {title} {\bibinfo {title} {Search for three-nucleon short-range correlations in light nuclei},\ }\href@noop {} {\bibfield  {journal} {\bibinfo  {journal} {Phys. Rev. C}\ }\textbf {\bibinfo {volume} {97}},\ \bibinfo {pages} {065204} (\bibinfo {year} {2018})}\BibitemShut {NoStop}%
\bibitem [{\citenamefont {Li}\ \emph {et~al.}(2022{\natexlab{b}})\citenamefont {Li} \emph {et~al.}}]{li2022revealing}%
  \BibitemOpen
  \bibfield  {author} {\bibinfo {author} {\bibfnamefont {S.}~\bibnamefont {Li}} \emph {et~al.},\ }\bibfield  {title} {\bibinfo {title} {Revealing the short-range structure of the mirror nuclei {$^3$H and $^3$He}},\ }\href@noop {} {\bibfield  {journal} {\bibinfo  {journal} {Nature}\ }\textbf {\bibinfo {volume} {609}},\ \bibinfo {pages} {41} (\bibinfo {year} {2022}{\natexlab{b}})}\BibitemShut {NoStop}%
\bibitem [{\citenamefont {Hen}\ \emph {et~al.}(2026)\citenamefont {Hen}, \citenamefont {Szumila-Vance},\ and\ \citenamefont {Weinstein}}]{Hen:2026hyi}%
  \BibitemOpen
  \bibfield  {author} {\bibinfo {author} {\bibfnamefont {O.}~\bibnamefont {Hen}}, \bibinfo {author} {\bibfnamefont {H.}~\bibnamefont {Szumila-Vance}},\ and\ \bibinfo {author} {\bibfnamefont {L.}~\bibnamefont {Weinstein}},\ }\bibfield  {title} {\bibinfo {title} {{Short-range correlations in nuclei}},\ }\href@noop {} {\  (\bibinfo {year} {2026})},\ \Eprint {https://arxiv.org/abs/2605.12626} {arXiv:2605.12626 [nucl-ex]} \BibitemShut {NoStop}%
\bibitem [{\citenamefont {Walet}\ \emph {et~al.}(2023)\citenamefont {Walet} \emph {et~al.}}]{Walet:2023myk}%
  \BibitemOpen
  \bibfield  {author} {\bibinfo {author} {\bibfnamefont {N.~R.}\ \bibnamefont {Walet}} \emph {et~al.},\ }\bibfield  {title} {\bibinfo {title} {{Calculation of Dynamical Response Functions Using a Bound-State Method}},\ }\href {https://doi.org/10.1007/s00601-023-01824-x} {\bibfield  {journal} {\bibinfo  {journal} {Few Body Syst.}\ }\textbf {\bibinfo {volume} {64}},\ \bibinfo {pages} {56} (\bibinfo {year} {2023})}\BibitemShut {NoStop}%
\bibitem [{\citenamefont {Walet}\ \emph {et~al.}(2026)\citenamefont {Walet} \emph {et~al.}}]{Walet:inprep}%
  \BibitemOpen
  \bibfield  {author} {\bibinfo {author} {\bibfnamefont {N.~R.}\ \bibnamefont {Walet}} \emph {et~al.}} (\bibinfo {year} {2026}),\ \bibinfo {note} {in preparation}\BibitemShut {NoStop}%
\bibitem [{\citenamefont {Hildebrandt}\ \emph {et~al.}(2010)\citenamefont {Hildebrandt}, \citenamefont {Griesshammer},\ and\ \citenamefont {Hemmert}}]{Hildebrandt:2010cqd}%
  \BibitemOpen
  \bibfield  {author} {\bibinfo {author} {\bibfnamefont {R.~P.}\ \bibnamefont {Hildebrandt}}, \bibinfo {author} {\bibfnamefont {H.~W.}\ \bibnamefont {Griesshammer}},\ and\ \bibinfo {author} {\bibfnamefont {T.~R.}\ \bibnamefont {Hemmert}},\ }\bibfield  {title} {\bibinfo {title} {{Nucleon polarizabilities from deuteron Compton scattering within a Green's-function hybrid approach}},\ }\href {https://doi.org/10.1140/epja/i2010-11024-y} {\bibfield  {journal} {\bibinfo  {journal} {Eur. Phys. J. A}\ }\textbf {\bibinfo {volume} {46}},\ \bibinfo {pages} {111} (\bibinfo {year} {2010})}\BibitemShut {NoStop}%
\bibitem [{\citenamefont {Lewis}(2025)}]{lewis2025nucleon}%
  \BibitemOpen
  \bibfield  {author} {\bibinfo {author} {\bibfnamefont {M.}~\bibnamefont {Lewis}},\ }\bibfield  {title} {\bibinfo {title} {{Nucleon Polarizabilities and Elastic Compton Scattering on {$^4$He} at 87 MeV and 100 MeV}},\ }\href@noop {} {\bibfield  {journal} {\bibinfo  {journal} {Bulletin of the American Physical Society}\ } (\bibinfo {year} {2025})}\BibitemShut {NoStop}%
\end{thebibliography}%

\onecolumngrid 
\section{End Matter}
\vspace{0.5cm}
\begin{center}
\includegraphics[width=\linewidth]{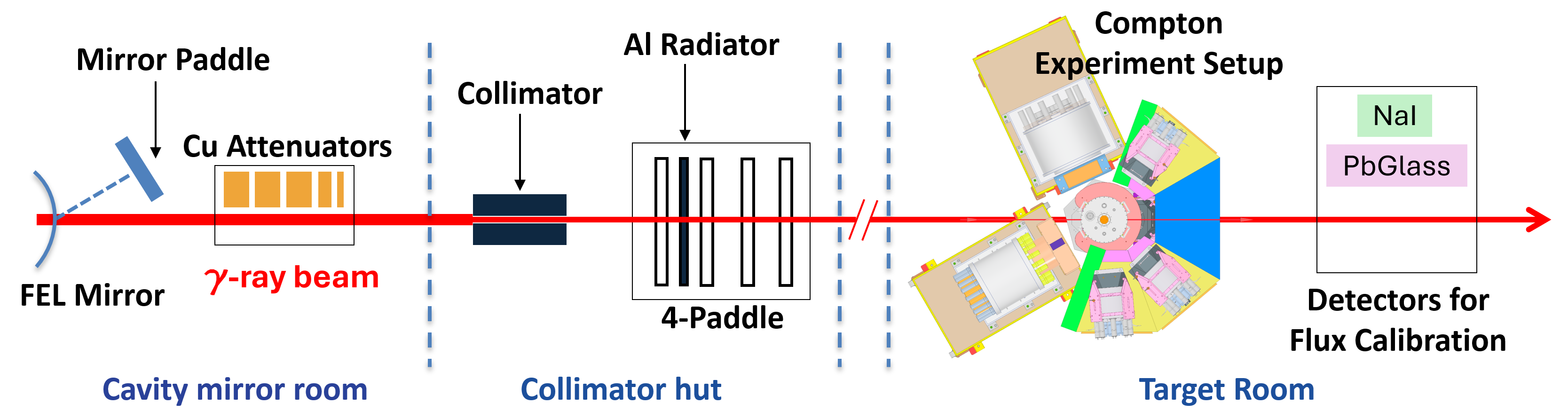}
\captionof{figure}{Configuration of the experiment, downstream of the HI$\gamma$S beamline during the production run (top view, not to scale). } 
\label{HIGS_production_config}
\end{center}
\vspace{0.5cm}
\twocolumngrid 

The collimated photon beam intensity was monitored using a newly built low-efficiency 4-paddle system, consisting of four plastic paddle scintillator detectors and one 2-mm-thick aluminum plate~\cite{em2026,jz2026}. This system was calibrated using two independent detectors downstream in the target room: a primary NaI(Tl) scintillator and a reference lead-glass (PbGlass) Cherenkov counter. The 4-paddle system was periodically calibrated against absolute photon counts measured with the NaI(Tl) detector. During these calibrations, copper attenuators of varying thicknesses, located downstream of the FEL mirror, were placed in the beamline to reduce the photon intensity to levels that the target-room detectors could reliably measure. The efficiency of the 4-paddle system was extracted from these measurements, allowing the unattenuated on-target beam intensity to be determined after applying small corrections for pileup~\cite{em2026,jz2026}.


\end{document}